\documentclass[aps,prd,superscriptaddress,preprint,tightenlines,nofootinbib,showpacs,floatfix]{revtex4}

\usepackage{epsfig}
\usepackage{graphicx}% Include figure files
\usepackage{graphics}
\usepackage{dcolumn}% Align table columns on decimal point
\usepackage{bm}% bold math
\usepackage{afterpage}

\newcommand{\etal}{{\it et al.}}

\begin{document}
%primary authors: Liming Zhang and Sheldon Stone

\preprint{\tighten\vbox{
 %                     \hbox{\hfil Sept. 17, 2009}
                      \hbox{\hfil CLNS 09/2061}
                      \hbox{\hfil CLEO 09-14}
}}
%\vspace*{2cm}
%title
\title{\boldmath Measurement of the Pseudoscalar Decay Constant $f_{D_s}$ Using $D_s^+\to\tau^+\nu$, $\tau^+\to\rho^+\overline{\nu}$ Decays}

\author{P.~Naik}
\author{J.~Rademacker}
\affiliation{University of Bristol, Bristol BS8 1TL, UK}
\author{D.~M.~Asner}
\author{K.~W.~Edwards}
\author{K.~Randrianarivony}
\author{J.~Reed}
\author{A.~N.~Robichaud}
\author{G.~Tatishvili}
\author{E.~J.~White}
\affiliation{Carleton University, Ottawa, Ontario, Canada K1S 5B6}
\author{R.~A.~Briere}
\author{H.~Vogel}
\affiliation{Carnegie Mellon University, Pittsburgh, Pennsylvania 15213, USA}
\author{P.~U.~E.~Onyisi}
\author{J.~L.~Rosner}
\affiliation{University of Chicago, Chicago, Illinois 60637, USA}
\author{J.~P.~Alexander}
\author{D.~G.~Cassel}
\author{R.~Ehrlich}
\author{L.~Fields}
\author{L.~Gibbons}
\author{S.~W.~Gray}
\author{D.~L.~Hartill}
\author{B.~K.~Heltsley}
\author{J.~M.~Hunt}
\author{D.~L.~Kreinick}
\author{V.~E.~Kuznetsov}
\author{J.~Ledoux}
\author{J.~R.~Patterson}
\author{D.~Peterson}
\author{D.~Riley}
\author{A.~Ryd}
\author{A.~J.~Sadoff}
\author{X.~Shi}
\author{S.~Stroiney}
\author{W.~M.~Sun}
\affiliation{Cornell University, Ithaca, New York 14853, USA}
\author{J.~Yelton}
\affiliation{University of Florida, Gainesville, Florida 32611, USA}
\author{P.~Rubin}
\affiliation{George Mason University, Fairfax, Virginia 22030, USA}
\author{N.~Lowrey}
\author{S.~Mehrabyan}
\author{M.~Selen}
\author{J.~Wiss}
\affiliation{University of Illinois, Urbana-Champaign, Illinois 61801, USA}
\author{M.~Kornicer}
\author{R.~E.~Mitchell}
\author{M.~R.~Shepherd}
\author{C.~M.~Tarbert}
\affiliation{Indiana University, Bloomington, Indiana 47405, USA }
\author{D.~Besson}
\affiliation{University of Kansas, Lawrence, Kansas 66045, USA}
\author{T.~K.~Pedlar}
\author{J.~Xavier}
\affiliation{Luther College, Decorah, Iowa 52101, USA}
\author{D.~Cronin-Hennessy}
\author{K.~Y.~Gao}
\author{J.~Hietala}
\author{R.~Poling}
\author{P.~Zweber}
\affiliation{University of Minnesota, Minneapolis, Minnesota 55455, USA}
\author{S.~Dobbs}
\author{Z.~Metreveli}
\author{K.~K.~Seth}
\author{B.~J.~Y.~Tan}
\author{A.~Tomaradze}
\affiliation{Northwestern University, Evanston, Illinois 60208, USA}
\author{S.~Brisbane}
\author{J.~Libby}
\author{L.~Martin}
\author{A.~Powell}
\author{P.~Spradlin}
\author{G.~Wilkinson}
\affiliation{University of Oxford, Oxford OX1 3RH, UK}
\author{H.~Mendez}
\affiliation{University of Puerto Rico, Mayaguez, Puerto Rico 00681}
\author{J.~Y.~Ge}
\author{D.~H.~Miller}
\author{I.~P.~J.~Shipsey}
\author{B.~Xin}
\affiliation{Purdue University, West Lafayette, Indiana 47907, USA}
\author{G.~S.~Adams}
\author{D.~Hu}
\author{B.~Moziak}
\author{J.~Napolitano}
\affiliation{Rensselaer Polytechnic Institute, Troy, New York 12180, USA}
\author{K.~M.~Ecklund}
\affiliation{Rice University, Houston, Texas 77005, USA}
\author{J.~Insler}
\author{H.~Muramatsu}
\author{C.~S.~Park}
\author{E.~H.~Thorndike}
\author{F.~Yang}
\affiliation{University of Rochester, Rochester, New York 14627, USA}
\author{S.~Ricciardi}
\affiliation{STFC Rutherford Appleton Laboratory, Chilton, Didcot, Oxfordshire, OX11 0QX, UK}
\author{C.~Thomas}
\affiliation{University of Oxford, Oxford OX1 3RH, UK}
\affiliation{STFC Rutherford Appleton Laboratory, Chilton, Didcot, Oxfordshire, OX11 0QX, UK}
\author{M.~Artuso}
\author{S.~Blusk}
\author{S.~Khalil}
\author{R.~Mountain}
\author{T.~Skwarnicki}
\author{S.~Stone}
\author{J.~C.~Wang}
\author{L.~M.~Zhang}
\affiliation{Syracuse University, Syracuse, New York 13244, USA}
\author{G.~Bonvicini}
\author{D.~Cinabro}
\author{A.~Lincoln}
\author{M.~J.~Smith}
\author{P.~Zhou}
\author{J.~Zhu}
\affiliation{Wayne State University, Detroit, Michigan 48202, USA}
\collaboration{CLEO Collaboration}
\noaffiliation

\date{October 18, 2009}

\begin{abstract}
Analyzing 600 pb$^{-1}$ of $e^+e^-$ collisions at 4170 MeV center-of-mass energy with the CLEO-c detector, we measure the branching fraction ${\cal B}(D_s^+\to \tau^+\nu)=(5.52\pm 0.57\pm 0.21)$\% using the $\tau^+\to\rho^+\overline{\nu}$ decay mode. Combining with other CLEO measurements of ${\cal B}(D_s^+\to \tau^+\nu)$ we determine the pseudoscalar decay constant $f_{D_s}=(259.7\pm 7.8\pm 3.4$) MeV consistent with the value obtained from our $D_s^+\to \mu^+\nu$ measurement of
($257.6\pm 10.3\pm 4.3$) MeV. Combining these measurements we find a value of $f_{D_s}=(259.0 \pm  6.2\pm 3.0$) MeV, that differs from the most accurate prediction based on unquenched lattice gauge theory of ($241\pm 3)$ MeV by 2.4 standard deviations. We also present the first measurements of
${\cal B}(D_s^+\to K^0\pi^+\pi^0)=(1.00\pm0.18\pm 0.04)$\%, and
${\cal B}(D_s^+\to\pi^+\pi^0\pi^0)=(0.65\pm0.13\pm 0.03)$\%, and measure a new value for
${\cal B}(D_s^+\to\eta\rho^+)=(8.9\pm0.6\pm0.5)$\%.

\end{abstract}

\maketitle \tighten
%\tableofcontents

%\newpage

\section{Introduction}

The purely leptonic decay of the $D_s^+$ meson occurs in the Standard Model (SM)
via the annihilation of the constituent charm quark with the constituent anti-strange
quark into a virtual $W^+$ boson that subsequently materializes as a lepton-antineutrino pair.
The SM decay rate is given by
\cite{Formula1}
\begin{equation}
\Gamma(D_s^+\to \ell^+\nu) = {{G_F^2}\over
8\pi}f_{D_s}^2m_{\ell}^2M_{D_s^+} \left(1-{m_{\ell}^2\over
M_{D_s^+}^2}\right)^2 \left|V_{cs}\right|^2~~~, \label{eq:equ_rate}
\end{equation}
where $M_{D_s^+}$ is the $D_s^+$ mass, $m_{\ell}$ is the mass of the
charged final state lepton, $G_F$ is the Fermi coupling constant,
 $|V_{cs}|$ is a Cabibbo-Kobayashi-Maskawa matrix element, and $f_{D_s}$ is the ``decay constant," a
parameter related to the overlap of the heavy and light quark
wave-functions at zero spatial separation.

Measurements of $D_s^+\to\mu^+\nu$ and $D_s^+\to\tau^+\nu$
have been made with increasing precision  recently,
and a disagreement  has emerged
between the theoretical value of $f_{D_s}$ computed by Follana \etal~\cite{Lat:Foll}, and  the
average of these measurements \cite{Rosner-Stone,PDG}.  It has been
pointed out by  Akeroyd and  Mahmoudi \cite{AkeroydM} that physics beyond the SM could
contribute differently to $\mu^+\nu$ and $\tau^+\nu$ final states, so increased precision on
each of these is important.

We report here on a new measurement of ${\cal {B}}(D_s^+\to\tau^+\nu)$ using the $\tau^+\to \rho^+\overline{\nu}$
decay mode. Previously CLEO has reported on this rate using the $\tau^+\to \pi^+\overline{\nu}$ and $\tau^+\to e^+\nu\overline{\nu}$ modes.

\section{Experimental Method}
\subsection{\boldmath Selection of $D_s$ Candidates}

The CLEO-c detector \cite{CLEODR} is equipped to measure the momenta
and directions of charged particles, identify them using specific
ionization ($dE/dx$) and Cherenkov light (RICH) \cite{RICH}, detect
photons and determine their directions and energies.

In this study we use 600 pb$^{-1}$ of data produced in $e^+e^-$
collisions using the Cornell Electron Storage Ring (CESR) and
recorded near a center-of-mass energy ($E_{\rm CM}$) of 4.170 GeV.
At this energy the $e^+e^-$ annihilation cross-section into
$D_s^-D_s^{*+}$ + $D_s^{*-}D_s^{+}$ is approximately 1~nb \cite{poling}.

\begin{figure}[hbt]
\centering
\includegraphics[width=6in]{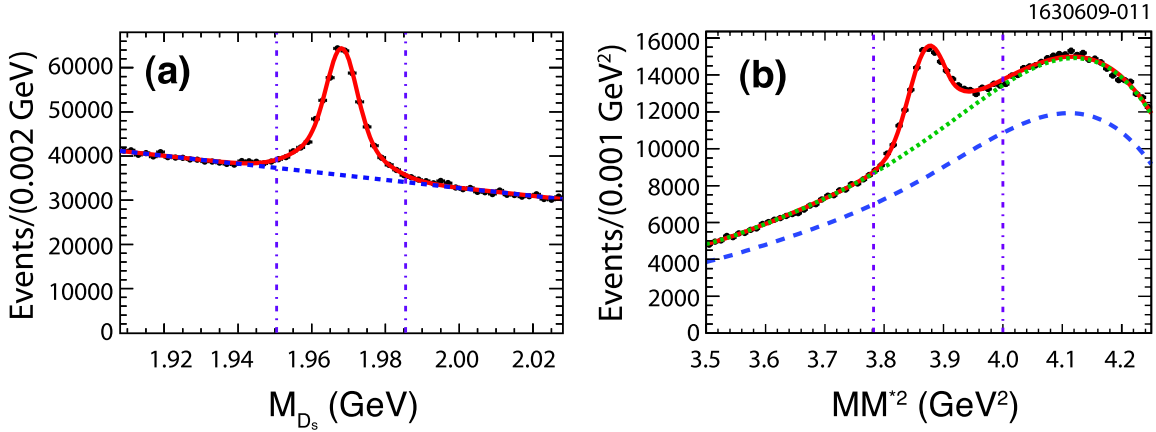}
\vspace{2mm}
\caption{(a) Invariant mass of $D_s^-$ candidates
summed over all decay modes and fit to a two-Gaussian signal shape
plus a straight line for the background. The vertical dot-dashed
lines indicate the $\pm$17.5 MeV definition of the signal region.  No MM$^{*2}$ cut has been
applied. (b)
The MM$^{*2}$ distribution summed over all modes. The
curves are fits to the number of signal events using the Crystal Ball  function
and two 5th order
Chebyshev background functions; the dashed curve shows the
background from fake $D_s^-$ tags, while the dotted curve in (b) shows
the sum of the backgrounds from multiple photon combinations and
fake $D_s^-$ tags. The vertical dashed lines show
the region of events selected for further analysis.
 } \label{mass-mm2-all}
\end{figure}

In this analysis we fully reconstruct a sample of $D_s^-$
in nine ``tag" modes and then find candidate $\rho^+\to\pi^+\pi^0$
decays in this sample. (Mention of any specific decay implies
the use of its charge-conjugate as well.) The tag selection is
identical to that used in our $D_s^+\to \mu^+\nu$ paper that
can be consulted for details \cite{Dstomunu}.
Briefly, we select candidates using their invariant masses. We require that the candidate energies are consistent with those expected of a $D_s$ or $D_s^*$ in $D_sD_s^*$ events.
%Briefly, we select
%candidates using their invariant mass, ensuring that their energies
%are consistent with $D_sD_s^*$ production.
The invariant mass distribution for
all tag candidates is shown in Fig.~\ref{mass-mm2-all}(a). Then we detect
an additional photon candidate from the $D_s^{*}$ decay, and construct
\begin{equation}
\label{eq:mmss} {\rm MM}^{*2}=\left(E_{\rm
CM}-E_{D_s}-E_{\gamma}\right)^2- \left({\bf p}_{\rm
CM}-{\bf p}_{D_s}-{\bf p}_{\gamma}\right)^2,
\end{equation}
where $E_{\rm CM}$ (${\bf p}_{\rm CM}$) is the
center-of-mass energy (momentum), $E_{D_s}$
(${\bf p}_{D_s}$) is the energy (momentum) of the fully
reconstructed $D_s^-$ tag, and $E_{\gamma}$
(${\bf p}_{\gamma}$) is the energy (momentum) of the
additional photon. In performing this calculation we use a kinematic
fit that constrains the decay products of the $D_s^-$ to the known
$D_s$ mass and conserves overall momentum and energy. All photon
candidates in the event are tried, except for those that are decay
products of the $D_s^-$ tag candidate.
 Regardless of whether
or not the photon forms a $D_s^*$ with the tag, for real $D_s^*D_s$
events the missing mass-squared MM$^{*2}$, recoiling against the
photon and the $D_s^-$ tag should peak at the $D_s^{+}$
mass-squared.

The MM$^{*2}$ distributions for events in the $D_s^-$ invariant mass
signal region ($\pm$17.5 MeV from the $D_s$ mass) are shown in
Fig.~\ref{mass-mm2-all}(b). In order to find the number of tags used for
further analysis we perform a two-dimensional binned maximum liklihood fit of the MM$^{*2}$
distribution and the invariant mass distribution in the interval $\pm$60 MeV from the
$D_s$ mass and $3.50 < {\rm MM}^{*2} <4.25$~GeV$^2$. The background has two components, both
described by 5th order Chebyshev polynomials; the first comes from
the background under the invariant mass peak, defined by the
sidebands, and the second is due to multiple photon combinations. In
both cases we allow the parameters to float.

 We find a
total of 43859$\pm$936$\pm$877 events within the interval
$3.782 < {\rm MM}^{*2}<4.000$ GeV$^2$ and having an invariant mass within $\pm$17.5 MeV of the $D_s$ mass, where the first uncertainty is
statistical and the second is systematic.

\subsection{Signal Reconstruction}
\label{sec:bfqsq}
We select events with one and only one charged track with opposite sign of charge to the tag, that is positively identified as a pion. The event also must contain at least one $\pi^0\to\gamma\gamma$ candidate with an invariant mass divided by the error on the invariant mass (Pull) $<$ 3; if there is more than one such candidate we choose the one with the minimum Pull. Tracks or photons that are used as part of the $D_s^-$ tag are not considered.  Unfortunately, hadron tracks can and do interact in the detector material, and deposit additional energy in the electromagnetic calorimeter. Thus we do not reject events with more than one $\pi^0$ candidate.
Photon candidates must have an energy deposition in the calorimeter consistent with that expected for an electromagnetic shower and deposit more than 30 MeV in the barrel or more than 50 MeV in the endcap.
In principle for $D_s^+\to\tau^+\nu,~\tau^+\to\rho^+\overline{\nu}$ decays all the energy should be accounted for in the decay products of the tag and and the $\rho^+$. We sum up any energy in the calorimeter not matched with tag or the $\rho^+$ and call this parameter $E_{\rm extra}$.

We also compute the MM$^2$  as
\begin{equation}
\label{eq:mm2} {\rm MM}^2=\left(E_{\rm
CM}-E_{D_s}-E_{\gamma}-E_{\rho}\right)^2
           -\left({\bf p}_{\rm CM}-{\bf p}_{D_s}
           -{\bf p}_{\gamma}
           -{\bf p}_{\rho}\right)^2,
\end{equation}
where $E_{\rho}$ (${\bf p}_{\rho}$) are the energy
(momentum) of the candidate $\rho^+$ and all other variables are the
same as defined in Eq.~(\ref{eq:mmss}).  While the MM$^2$ does not peak at zero, because there are two missing neutrinos,
it is still a useful variable as two-body $D_s^+$ decay backgrounds will peak; {\it e.g.}, $D_s^+\to\rho^+\eta$ peaks at the $\eta$ mass-squared.

We proceed by defining a $\pi^+\pi^0$ mass window consistent with the $\rho^+$ mass. Fortunately, the mass distribution and the branching ratio for $\tau^+\to\rho^+\overline{\nu}$ decays are well measured \cite{Belle-tau-rhonu}. We select events within $\pm$250 MeV of the $\rho^+$ mass. This selection is chosen to maximize efficiency while still not including too much background. This mass selection is 89.3\% efficient for $\tau^+\to\rho^+\overline{\nu}$ events.

\begin{figure}[hbt]
\centering
\includegraphics[width=6in]{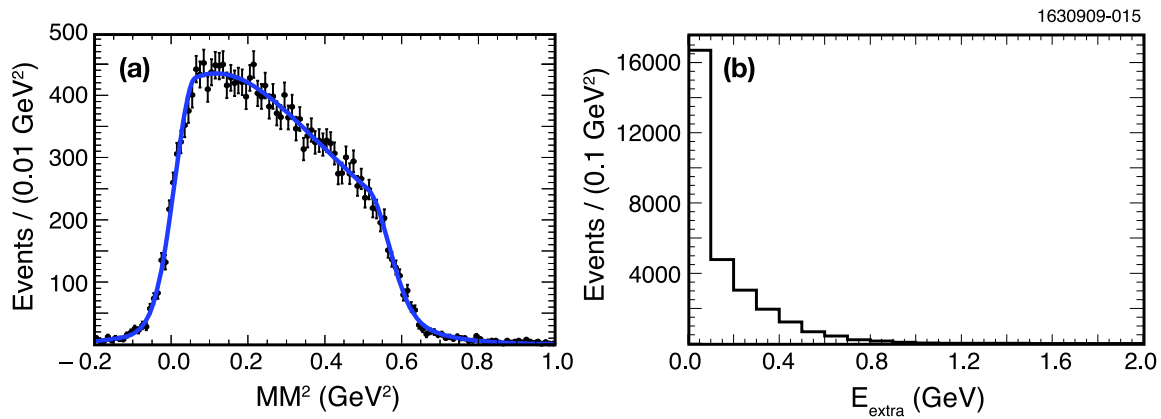}
\vspace{2mm}
\caption{Monte Carlo generated distributions for signal $D_s^+\to\tau^+\nu$, $\tau^+\to\rho^+\overline{\nu}$ (a) MM$^2$, and  (b) $E_{\rm extra}$. The curve in (a) is a fit to the data points with the sum of two bifurcated-Gaussian functions.
 } \label{mm2_sig}
\end{figure}

The expected MM$^2$ and $E_{\rm extra}$ distributions from signal Monte Carlo simulation of $D_s^-$ tag and $D_s^+\to\tau^+\nu$ events are shown in Fig.~\ref{mm2_sig}. Here we have included the above mentioned selection window on the $\pi^+\pi^0$ mass. The MM$^2$ signal shape is fit to a function that is the sum of two bifurcated-Gaussian functions.
(A bifurcated-Gaussian shape has different widths below and above the mean.)

\subsection{Background Expectations}
There are two general sources of background expected arising from either combinatoric background in the reconstructed $D_s^-$ tag sample, or specific decay modes of the $D_s^+$. The former are determined by using sidebands of the candidate $D_s^-$ invariant mass distribution. The latter could arise from modes involving $\rho^+$ decays such as $\eta\rho^+$, but could also come from any mode that includes a $\pi^+$ and a $\pi^0$, or a $\pi^+$ and extra energy that is called a $\pi^0$.
Previous studies have shown that requiring the $\pi^+$ candidate to project to the primary event vertex eliminates fake charged tracks as a background source \cite{oldDp}.
Our first look at the background from $D_s^+$ decays uses Monte Carlo simulation. The background MM$^2$ and $E_{\rm extra}$ distributions are shown in Fig.~\ref{mm2_Ee-gen}. The specific background modes are enumerated in Appendix A for three different intervals of extra energy,  $E_{\rm extra}$ $<$ 0.1 GeV, which we expect is dominated by signal, $0.1~<E_{\rm extra}<$ 0.2 GeV, which we expect has similar amounts of signal and background, and 0.8 GeV$\,<E_{\rm extra}$, where the signal is absent. We separate into these three intervals in order to test our understanding of the background.

\begin{figure}[hbt]
\centering
\includegraphics[width=6in]{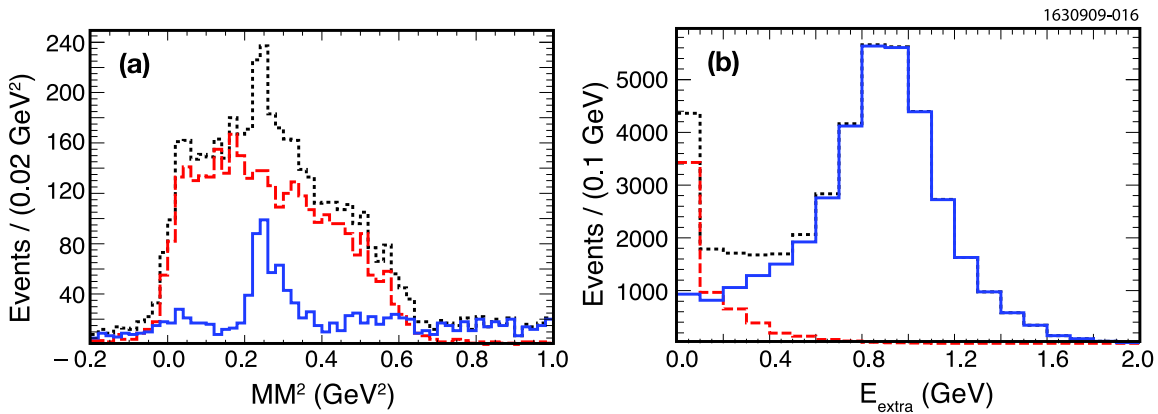}
\vspace{2mm}
\caption{Distributions for generic Monte Carlo backgrounds from $D_s^+$ decays (solid line) in (a) MM$^2$ for $E_{\rm extra}~<$~0.1 GeV, and  (b) $E_{\rm extra}$ for $-0.2<{\rm MM}^2<0.6$ GeV$^2$. The expected signal (dashed line) and total (dotted line) are also shown. The distributions are normalized to the Monte Carlo expectations.
 } \label{mm2_Ee-gen}
\end{figure}

Three final states cause narrow peaks in MM$^2$:  (i)  $K^0\pi^+\pi^0$ peaks at the $K^0$ mass-squared and has not been previously measured, (ii) $\pi^+\pi^0\pi^0$  peaks at the $\pi^0$ mass-squared and is also unmeasured, and (iii) $\eta\rho^+$ peaks at the $\eta$ mass-squared, and is poorly determined. In order to properly treat the background we measure the branching fractions of these modes using a double tag technique as described in Appendix B.

\subsection{\boldmath MM$^2$ Resolution}
\label{subsec:MM2}
\begin{figure}[hbt]
\centering
\includegraphics[width=6in]{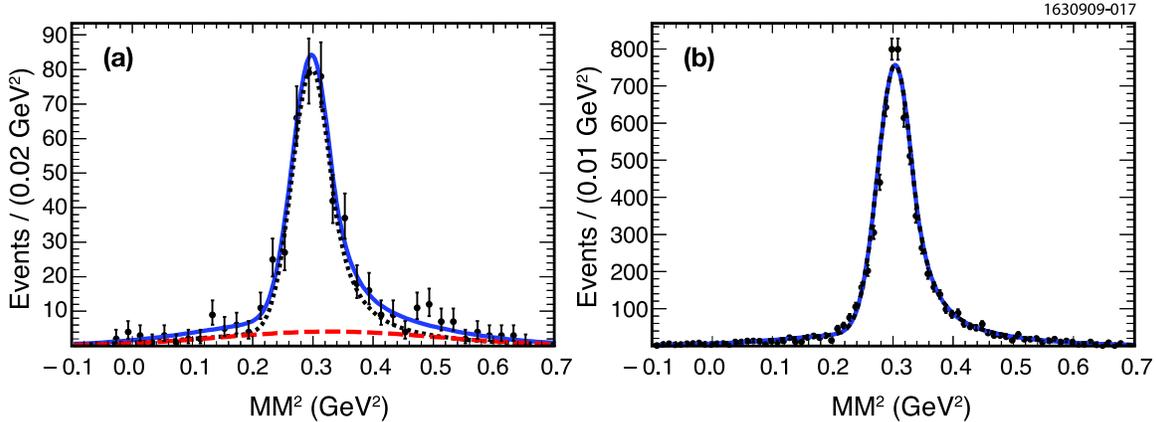}
\vspace{2mm}
\caption{ MM$^2$ spectrum of $\eta\pi^+\pi^0$, where $\eta\to\gamma\gamma$ was detected, but ignored in the calculation for both (a) data and (b) Monte Carlo events. The dotted line in (a) is the signal and the dashed line the background determined from the fit to the $D_s^-$ candidate invariant mass sidebands. The signal shapes are the sum of a Crystal Ball  and Gaussian functions (see text).
 } \label{mm2_rhonu}
\end{figure}

While the MM$^2$ resolution of the signal is not an issue because the signal does not form a narrow peak, several of
the backgrounds do have narrow structures and it is necessary to model their shapes properly.  In Fig.~\ref{mm2_rhonu} we compare the MM$^2$ distribution for $D_s^+\to\eta\rho^+$ signal from the Monte Carlo simulation with the one found in the data where the $\eta\to\gamma\gamma$ decay was detected. We also require that $\gamma\gamma\pi^+\pi^0$ mass be between 1.85 and 2.10 GeV, in order to have a relatively clean sample but not distort the MM$^2$ shape. The MM$^2$ is computed using Eq.~\ref{eq:mm2} while ignoring the two photons in the $\eta$ decay.

In both the case of the Monte Carlo simulation and the data we fit the signal with the sum of Crystal Ball (CB)  \cite{CBL}  and Gaussian functions. We fix some fit parameters that we find from the Monte Carlo simulation including the ratio of the r.m.s.\ widths ($\sigma$'s) for  the Gaussian and CB functions, set to a value of 6, and the area of the Gaussian function with respect to the CB function to be 20\%. In the data fit we also include the background given by the $D_s^-$ invariant mass sidebands. The CB function parameters $\alpha$ and $N$ are taken from the Monte Carlo simulation as 1 and 4.5, respectively. We find
\begin{eqnarray}
\sigma_{\rm MC}&=&(0.0289\pm0.0006)~{\rm GeV}^2~{\rm for~the~Monte~Carlo~CB~function,}\\\nonumber
\sigma_{\rm Data}&=&(0.0320\pm0.0020)~{\rm GeV}^2~{\rm for~the~data~CB~function,}\\\nonumber
\sqrt{\sigma_{\rm Data}^2-\sigma_{\rm MC}^2}&=&(0.014\pm 0.005)~{\rm GeV}^2~.
\end{eqnarray}

We use the resolution as found in the data above when fitting the data for the $\eta\rho^+$ component and increase the width of the other narrow components.

\section{Signal Extraction}
We proceed by preforming a simultaneous fit to the $D_s^-$ invariant mass,
using a mass range within $\pm$70 MeV of the nominal mass and the MM$^2$. The procedure is similar to that used in Ref.~\cite{Dstomunu}.
We first check our procedures by fitting the data MM$^2$ distribution in the $E_{\rm extra}$ interval above 0.8 GeV, where
we have only background. We include the following decay modes as individual probability density functions (PDFs) in the fit: $K^0\pi^+\pi^0$, $\pi^+\pi^0\pi^0$, $\phi\pi^+$, $\eta\pi^+$, $\eta\rho^+$, $\eta'\pi^+$, $\eta'\pi^+\pi^0$, $\omega\pi^+\pi^0$. All narrow resonance structures are smeared by an additional r.m.s.\ resolution of 0.014 GeV$^2$, as determined by our $\eta\rho^+$ study (see section~\ref{subsec:MM2}).
 The other modes are lumped together into one other PDF. In the likelihood fit we add Gaussian constraints on the expected yields based on the known branching ratios and their errors.

\begin{figure}[htb]
\centering
\includegraphics[width=6in]{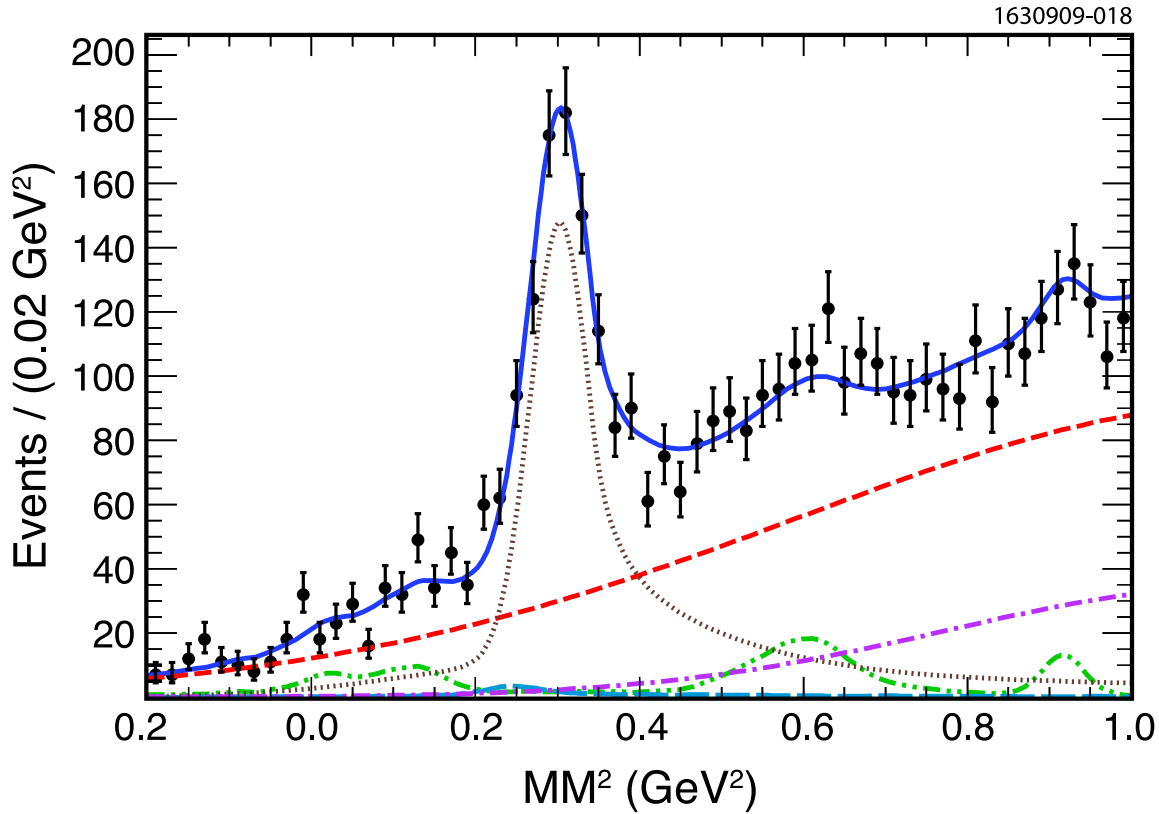}
\vspace{2mm}
\caption{Fit to the data (points)  for 0.8 GeV $<E_{\rm extra}$. The various components are   $\eta\rho^+$ (dotted), fake $D_s^-$ (dashed),  $K^0\pi^+\pi^0$ (long dash),  sum of  $\pi^+\pi^0\pi^0$, $\eta\pi^+$, $\omega\pi^+\pi^0$, $\phi\pi^+$, $\eta'\pi^+$, and $\eta'\pi^+\pi^0$  (dash-dot-dot), and other backgrounds (dashed-dot). The solid curve shows the total.
 } \label{fit_800MeV}
\end{figure}

The resulting fit to the data for 0.8 GeV $<~E_{\rm{extra}}$~is shown in Fig.~\ref{fit_800MeV}. The fake $D_s^-$ background has been accounted for by simultaneously fitting the sidebands in $D_s^-$ invariant mass. The two-body modes show evident peaks and are well described by the fit, demonstrating that our understanding of the backgrounds appears to be adequate.

\begin{figure}[hbt]
\centering
\includegraphics[width=6in]{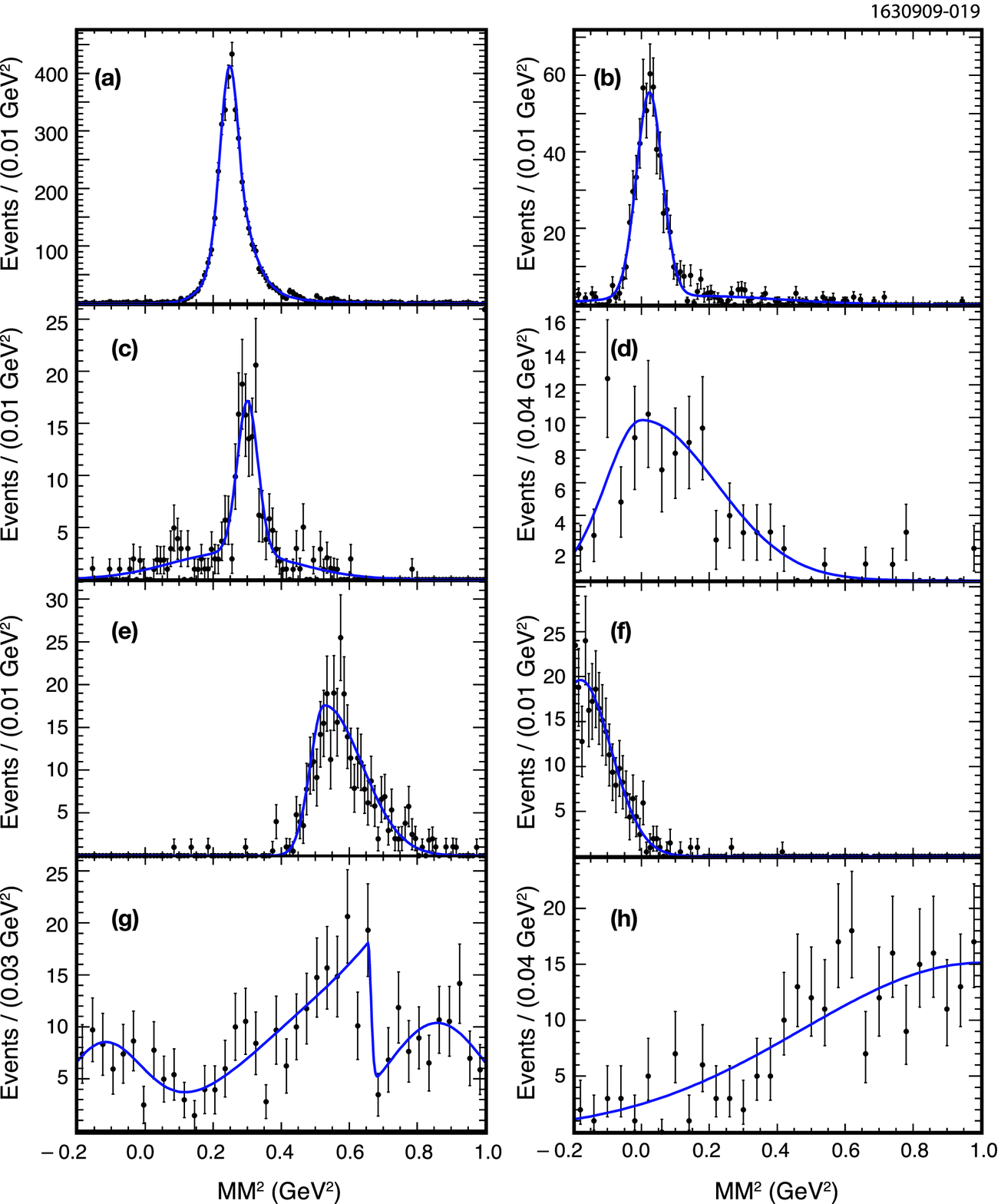}
\vspace{2mm}
\caption{Fits to Monte Carlo simulation for the individual background PDFs for $E_{\rm extra}<$ 0.1 GeV, for the modes (a) $K^0\pi^+\pi^0$, (b) $\pi^+\pi^0\pi^0$, (c) $\eta\rho^+$, (d) $\eta\pi^+$, (e) $\phi\pi^+$, (f) $\mu^+\nu$, (g) $D_s^+\to\tau^+\nu$, $\tau^+\to (\pi^++\pi^+\pi^0\pi^0)\overline{\nu}$, (h) sum of the other small modes.
 } \label{pdf_100}
\end{figure}

\begin{figure}[hbt]
\centering
\includegraphics[width=6in]{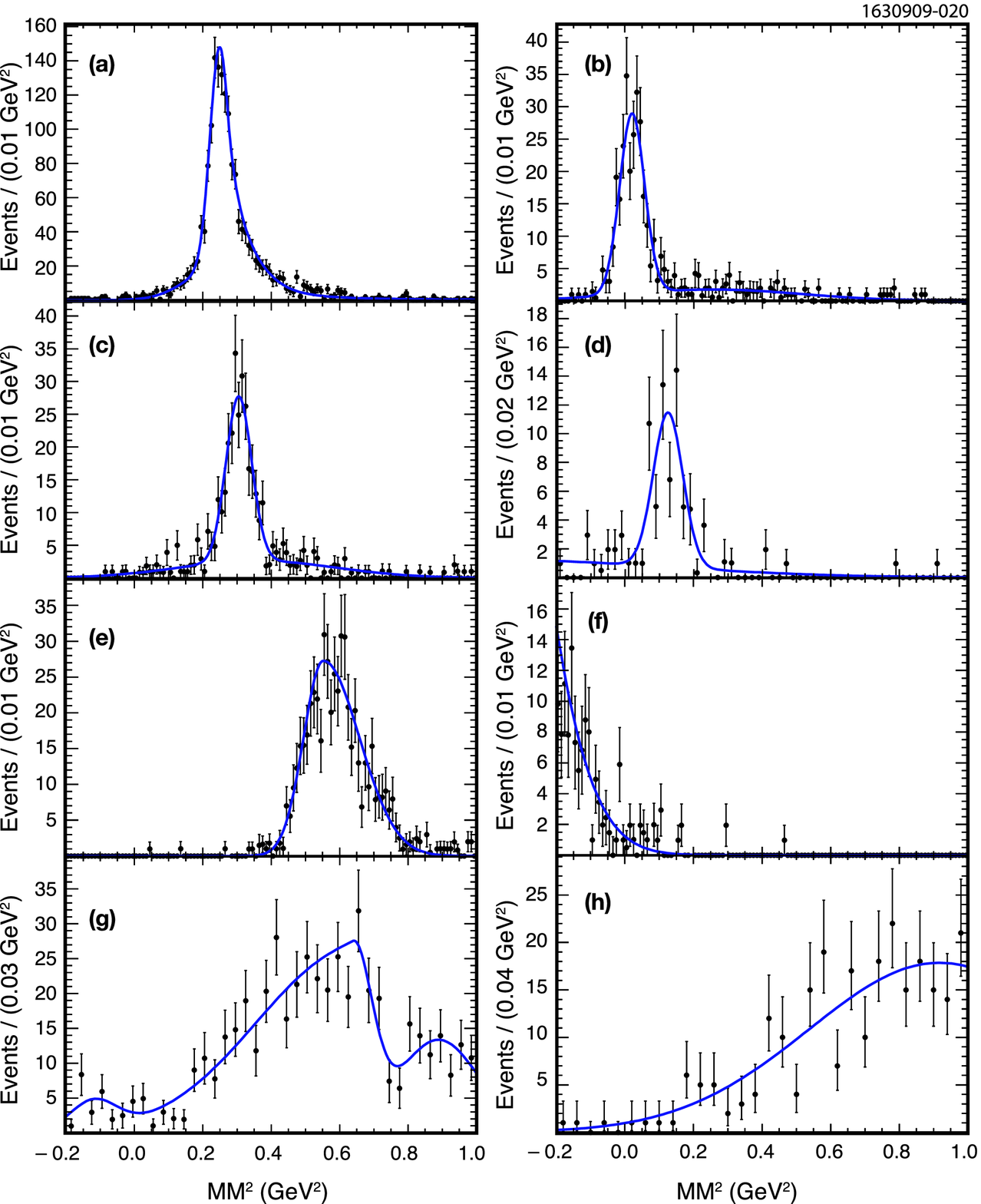}
\vspace{2mm}
\caption{Fits to Monte Carlo simulation  for the individual background PDFs for 0.1 $<E_{\rm extra}<$ 0.2 GeV, for the modes (a) $K^0\pi^+\pi^0$, (b) $\pi^+\pi^0\pi^0$, (c) $\eta\rho^+$, (d) $\eta\pi^+$, (e) $\phi\pi^+$, (f) $\mu^+\nu$, (g) $D_s^+\to\tau^+\nu$, $\tau^+\to (\pi^++\pi^+\pi^0\pi^0)\overline{\nu}$, (h) sum of the other small modes.
 } \label{pdf_200}
\end{figure}

\begin{figure}[hbt]
\centering
\includegraphics[width=6in]{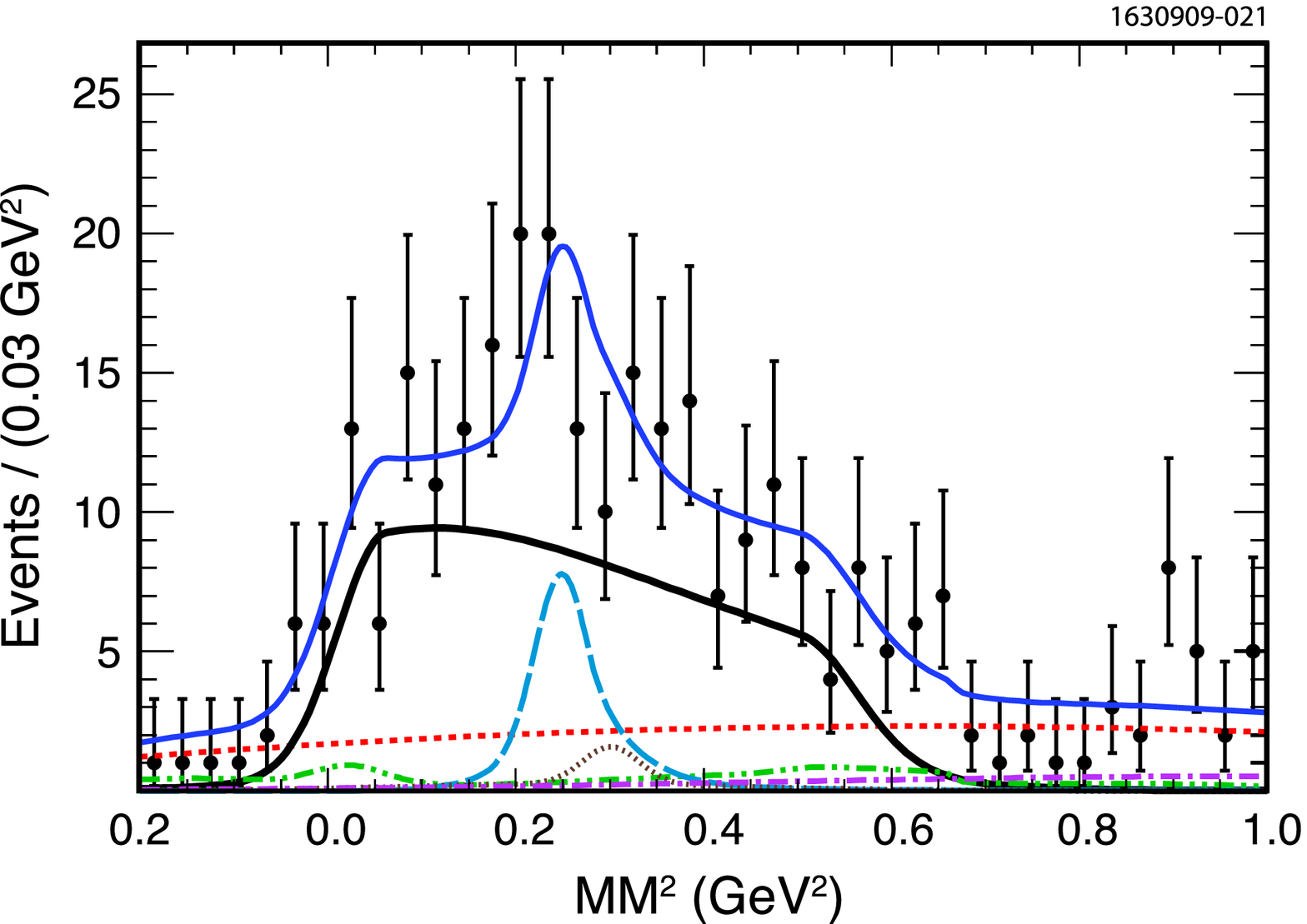}
\vspace{2mm}
\caption{Fit to the data (points)  for $E_{\rm extra}<$ 0.1 GeV. The various components are  signal (thick solid line), $\eta\rho^+$ (dotted), fake $D_s^-$ (dashed),  $K^0\pi^+\pi^0$ (long dash),  sum of  $\pi^+\pi^0\pi^0$, $\eta\pi^+$,  $\phi\pi^+$, $\tau^+\to(\pi^++\pi^+\pi^0\pi^0)\overline{\nu}$, $\mu^+\nu$, and $X\mu^+\nu$  (dash-dot-dot), and other backgrounds (dashed-dot). The thinner solid curve shows the total.
 } \label{fit_100MeV}
\end{figure}

\begin{figure}[hbt]
\centering
\includegraphics[width=6in]{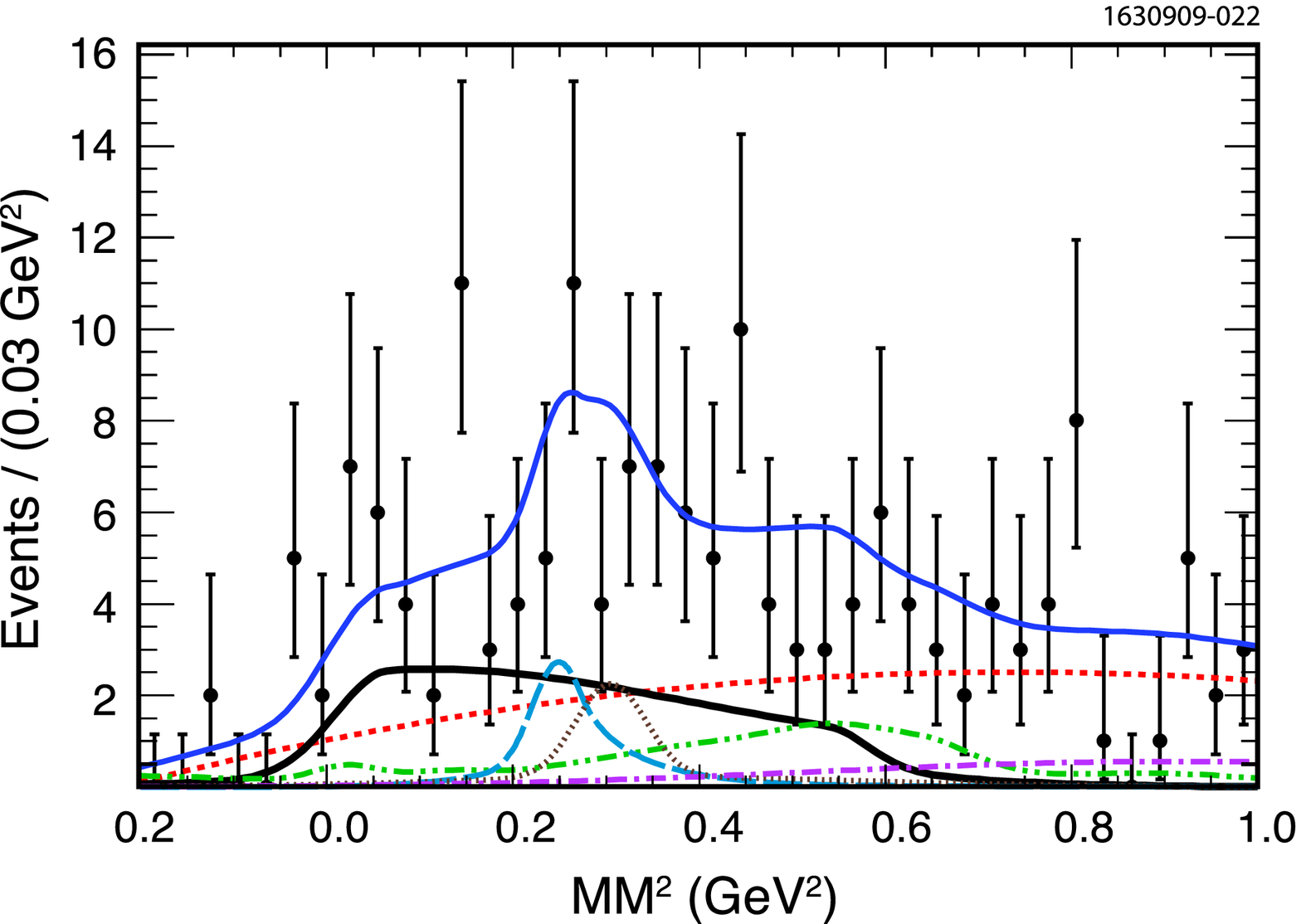}
\vspace{2mm}
\caption{Fit to the data (points)  for $0.1~<E_{\rm extra}<$ 0.2 GeV. The various components are  signal (thick solid line), $\eta\rho^+$ (dotted), fake $D_s^-$ (dashed),  $K^0\pi^+\pi^0$ (long dash),  sum of  $\pi^+\pi^0\pi^0$, $\eta\pi^+$,  $\phi\pi^+$, $\tau^+\to(\pi^++\pi^+\pi^0\pi^0)\overline{\nu}$, $\mu^+\nu$, and $X\mu^+\nu$  (dash-dot-dot), and other backgrounds (dashed-dot). The thinner solid curve shows the total.
 } \label{fit_100-200MeV}
\end{figure}

We next fit the two bins $E_{\rm extra}<0.1$ GeV and $0.1<E_{\rm extra}<0.2$ GeV separately.  The background PDFs are shown in Figs.~\ref{pdf_100} and \ref{pdf_200} for each interval. Note that they now include a separate PDF for $D_s^+\to\tau^+\nu$, where the $\tau^+$ decays into either $\pi^+\overline{\nu},~{\rm or}~\pi^+\pi^0\pi^0\overline{\nu}$.
The fit projections for both intervals are shown in Figs.~\ref{fit_100MeV} and \ref{fit_100-200MeV}.  For the first interval, $E_{\rm extra}<0.1$ GeV, the background level from all $D_s^+$ decays is about the same as that from fake $D_s^-$, and both are considerably smaller than the signal. For the second interval the signal and background levels are about equal. Table~\ref{tab:fits} summarizes the signal and background yields. (The notation  ``$\tau^+\to (\pi^++\pi^+\pi^0\pi^0)\overline{\nu}$" denotes the sum of two modes where $D_s^+\to\tau^+\nu$,  and the $\tau^+$ decays to either $\pi^+\overline{\nu}$ or $\pi^+\pi^0\pi^0\overline{\nu}$.)

\afterpage{\clearpage}

\begin{table}[htb]
\begin{center}
\caption{Signal and background yields from the fit in two $E_{\rm extra}$ intervals. We also list the measured or assumed background branching fractions and the r.m.s.\ error on the fit constraint resulting either from the branching fraction error or other considerations. The \# MC indicates the predicted background number of events for the assumed branching ratio input as the starting point of the fit, while \# Data gives the number determined by the fit.\label{tab:fits}}
\begin{tabular}{lccrrrr} \hline\hline
Component&${\cal B}$(\%) & Constraint&\multicolumn{2}{c}{$E_{\rm extra}<0.1$ GeV}&\multicolumn{2}{c}{$0.1<E_{\rm extra}<0.2$ GeV }\\
&& Error  (\%) &\# MC~ &\# Data~~~ & \# MC~ &\# Data~~~\\\hline
Signal &&&&155.2$\pm$16.5 & &43.7$\pm$11.3\\
$K^0\pi^+\pi^0$ &1.0$\pm$0.2 & 20 & 26.1&25.2$\pm$4.8 & 11.0 & 10.5$\pm$2.1\\
$\eta\rho^+$ & $8.9\pm 0.7$&4.2 & 7.1 &7.0$\pm$0.6 & 10.6 &10.5$\pm$0.9\\
$\pi^+\pi^0\pi^0$ & $0.65\pm 0.14$ &22 &2.8 & $2.8\pm 0.6$ & 1.5 & $1.6\pm 0.3$\\
$\tau^+\to (\pi^++\pi^+\pi^0\pi^0)\overline{\nu}$ &$1.14\pm 0.06$  &25${^\dagger}$& 8.5&$8.4\pm 2.1$&12.2&$10.9\pm 3.0$\\
$\mu^+\nu$ & $0.576\pm  0.045$ & 5.4 & 1.0 & $1.0\pm 0.1$& 0.48 &$0.5\pm 0.1$\\
$\eta\pi^+$ & $1.58\pm0.21$ & 13.3 & 0.9 & $0.9\pm 0.1$ & 0.9 & $0.9\pm 0.1$ \\
$\phi\pi^+$ & $4.35\pm 0.35$ & 8 & 1.7 &$1.7\pm 0.2$ & 2.8 & $2.8\pm 0.3$\\
$X\mu^+\nu$& 5.9 & 35$^{\ddagger}$ & 3.4 & $3.4\pm 1.2$ & 7.4 & $6.6\pm 2.6$\\
Other background & & $30^*$ & 11.5  &$11.4\pm 3.3$ & 11.8 &$10.5\pm 3.3$ \\
Fake $D_s^-$ background& &&&$81.8\pm 5.0$& & $74.8\pm 4.6$\\
\hline\hline
\multicolumn{7}{l}{$\dagger$ The error is  based on the uncertainties of the resonant substructure that can alter the efficiency.}\\
\multicolumn{7}{l}{$\ddagger$ We  assign a 35\% uncertainty based upon the error on  ${\cal{B}}(D_s^+\to X e^+ \nu)$.}\\
\multicolumn{7}{l}{$*$ We assign a 30\% uncertainty based on the sample size.}\\
\end{tabular}
\end{center}

\end{table}

Adding the signal yields in the two $E_{\rm extra}$ intervals, taking into account the efficiency for finding the $\rho^+$ in each interval, and dividing by the number of $D_s^-$ tags (43859$\pm$936$\pm$877) we find
\begin{equation}
{\cal B}(D_s^+\to\tau^+\nu)=(5.52\pm0.57\pm0.21)\%,
\end{equation}
where the first error is statistical and the second systematic. We will discuss the systematic errors in the next section.
In the smallest $E_{\rm extra}$ interval the branching fractions is $(5.48\pm 0.59)$\%, while in the higher interval it is $(5.65\pm 1.47)$\%. The numbers are consistent.  We note the data including the background components
is well-modeled in all three $E_{\rm extra}$ intervals, confirming our understanding of the background.

\subsection{Systematic Errors}

The sources of systematic errors in the branching fraction are listed in Table~\ref{tab:syserr}.
As we have let the branching fractions of the background components float in the fit by their known errors, there is no additional contribution from this source. The systematic error in the background is estimated using two different techniques.  First of all, if we remove the Gaussian constraint on the sum of the other small mode background fractions we observe a 1.1\% increase in the signal yield. Secondly, if we change the parameters of the background shape containing the sum of the other small modes the yield decreases by 0.5\%. A separate source of error is the efficiency on the detection of background events; if we change the $E_{\rm extra}$ efficiency and the $\pi^0$ efficiency by their errors, and thus change the background yields, we observe a combined error of $\pm$1.1\%.
An additional systematic error could arise from Cabibbo suppressed $\tau^+\to K^+\pi^0\overline{\nu}$ decays. The measured branching fraction for these decays is 1.6\% of that of $\pi^+\pi^0\overline{\nu}$. The combination of $dE/dx$ and RICH particle identification reduces the kaon yield by more than 95\%, resulting in a negligible $<0.1$\% contamination.

\begin{table}[htb]
\begin{center}
\caption{Systematic errors on determination of the branching fraction. \label{tab:syserr}}
\begin{tabular}{lc} \hline\hline
   Error Source & Size (\%) \\ \hline
Finding the $\pi^+$ track from the $\rho^+$ decay  &0.3 \\
Hadron identification & 1.0\\
Finding the $\pi^0$  from the $\rho^+$ decay  &1.3 \\
$E_{\rm extra} <$ 0.2 GeV \& $\pi^0$ efficiencies on background& 1.1\\
$E_{\rm extra} <$ 0.2 GeV signal efficiency & 2.0\\
Background modeling& 1.1\\
Number of tags& 2.0\\
Tag bias & 1.0\\
\hline
Total & 3.8\\
 \hline\hline
\end{tabular}
\end{center}
\end{table}

\begin{figure}[hbt]
\centering
\includegraphics[width=4in]{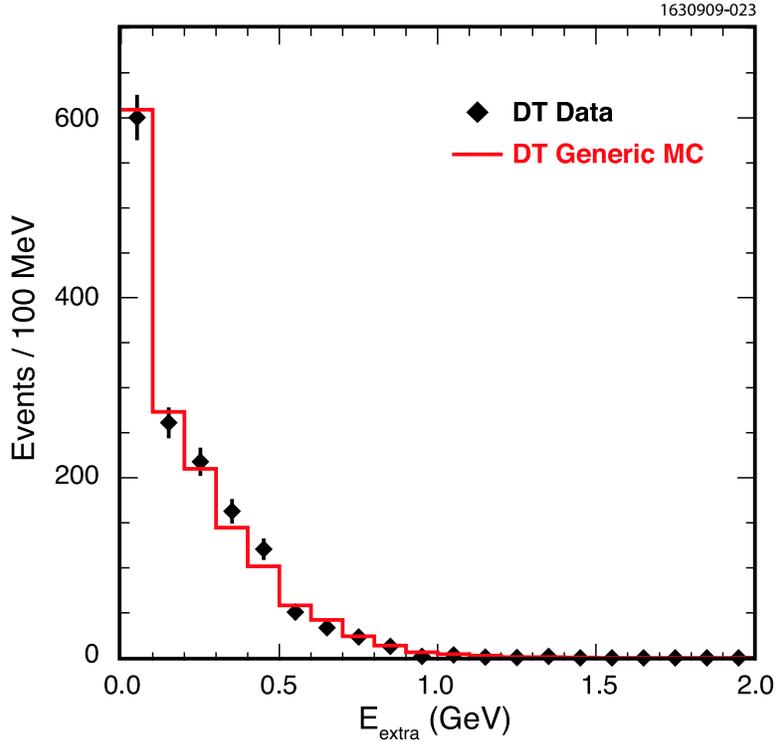}
\vspace{2mm}
\caption{The $E_{\rm extra}$ distributions from $e^+e^-\to D_s^*D_s$ events with both $D_s$ decays fully reconstructed (DT) for data (diamond) and Monte Carlo simulation (histogram).
 } \label{etot_doubletag}
\end{figure}

Since we are requiring that $E_{\rm extra}$ be below either 0.1 or 0.2 GeV, it is necessary to check this
efficiency in the data.
Our procedure is to use the fully reconstructed sample of $e^+e^-\to D_sD_s^*$ events selected the same way as described in Ref.~\cite{Dstomunu}. The $E_{\rm extra}$ distributions from Monte Carlo simulation and data for this sample are shown in Fig.~\ref{etot_doubletag}.  The agreement with the simulation is excellent. Table~\ref{tab:Eextra} gives the efficiency from  Monte Carlo simulation for our double tag sample and the efficiency measured in the data for specific ranges in $E_{\rm extra}$. The situation here corresponds to the extra energy deposited by two tags. We need to translate these numbers to the case of one tag plus a signal $\rho^+$. A cut at $E_{\rm extra}=0.3$ GeV in the double tag data corresponds to the same efficiency as a cut of 0.2 GeV in the single tag plus $\rho^+$ data. The difference between Monte Carlo simulation and data then is $(-1.2\pm 1.6)$\%, which implies a systematic error of $\pm$2.0\% in this efficiency.

\begin{table}[htb]
\begin{center}
\caption{The efficiencies ($\epsilon$) for Data and Monte Carlo simulation for different
requirements on $E_{\rm extra}$, and the corresponding fractional differences. \label{tab:Eextra}}
\begin{tabular}{cccc} \hline\hline
$E_{\rm extra}$ (GeV)& $\epsilon_{\rm Data}$(\%)& $\epsilon_{\rm MC}$(\%)& $\epsilon_{\rm Data}/\epsilon_{\rm MC}-1$ (\%)\\\hline
$<$0.1 & $40.24\pm 1.27$ & $40.81\pm 0.31$ &$-1.4\pm 3.2$\\
$<$0.2	&$57.75\pm1.28$&	$59.12\pm	0.31$&$-2.3\pm2.2$\\
$<$0.3	&$72.35\pm	1.16$&$73.21\pm0.28$	&$-1.2\pm	1.6$\\
$<$0.4	&$83.27	\pm0.97$	&$82.91\pm	0.24$	&~$0.4\pm1.2$\\
\hline\hline
\end{tabular}
\end{center}
\end{table}

We note that if we fix the background branching fractions to their nominal values, and refit the data, the statistical error in the $E_{\rm extra}<$ 0.1 GeV  bin decreases from 16.5 to 15.9 events, and in the 0.1 $<E_{\rm extra}<$ 0.2 GeV bin decreases from 11.3 to 11.1 events. Thus, our statistical error contains a significant component from the background estimates.

\subsection{\boldmath Cross Checks Using $\pi^+\pi^0$ Helicity and Mass Distributions}
\begin{figure}[hbtp]
\centering
\includegraphics[width=6in]{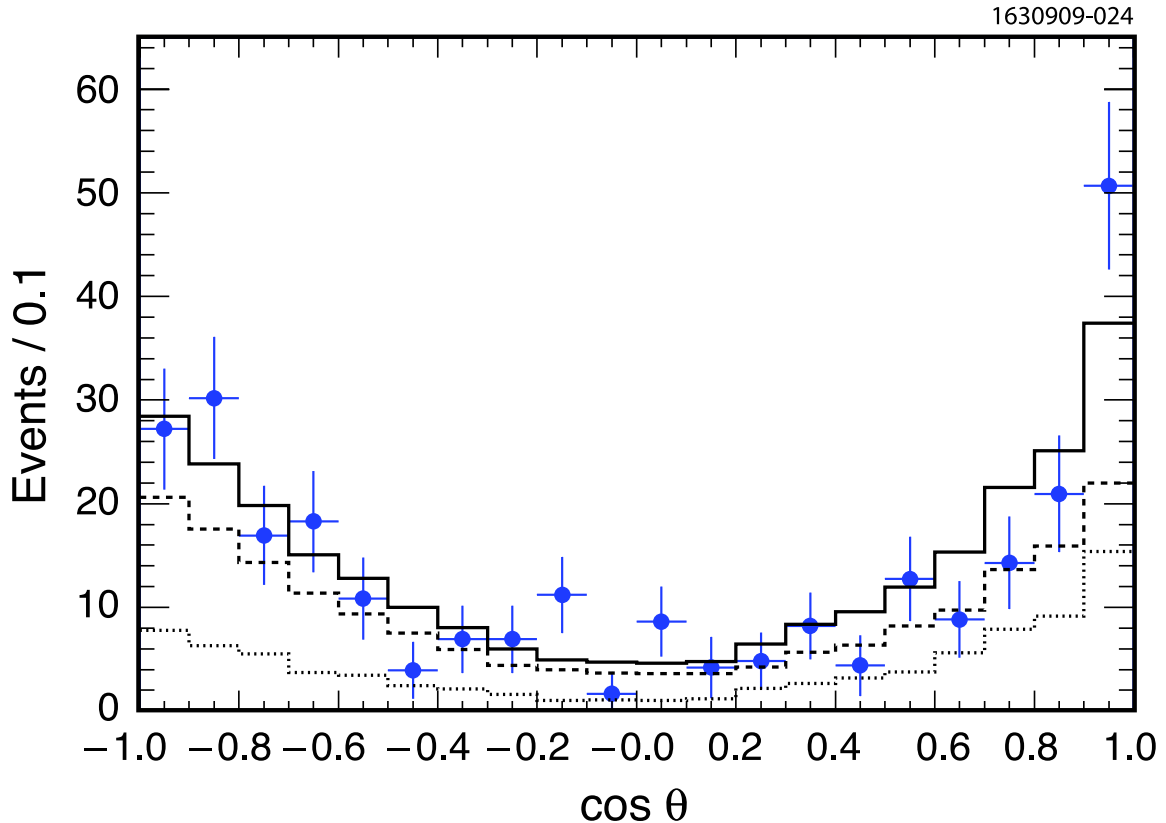}
\vspace{2mm}
\caption{Helicity distribution from the $\rho^+$ decay as measured in the laboratory frame.  The points with error bars are the sideband subtracted data. The dashed line represents the predicted signal shape and the dotted line the predicted background shape from real $D_s^+$ decays. The Monte Carlo predictions are normalized by the fitted yields to the data. The predicted total is given by the solid line. We require that  $E_{\rm extra}~<$ 0.2 GeV, and $ -0.05 ~< $MM$^2~ <$ 0.60 GeV$^2$.
 } \label{hel_rho}
\end{figure}

In principle the best way to view the $\rho^+$ polarization is to look at the angle $\theta$ of the $\pi^+$ with respect to the $\rho^+$ direction in the $\tau^+$ rest frame. Since we cannot reconstruct the $\tau^+$, we use the laboratory frame. We consider all events in the $E_{\rm extra}$ interval below 0.2 GeV and having $ -0.05 ~< $MM$^2~ <$ 0.60 GeV$^2$.  Fig.~\ref{hel_rho} shows $\cos\theta$ from the data with sidebands subtracted compared with the sum of expected signal and backgrounds from Monte Carlo simulation, normalized by yields from the data fit.  The predicted total is in good agreement with the shape and data yield.

\begin{figure}[hbtp]
\centering
\includegraphics[width=5.5in]{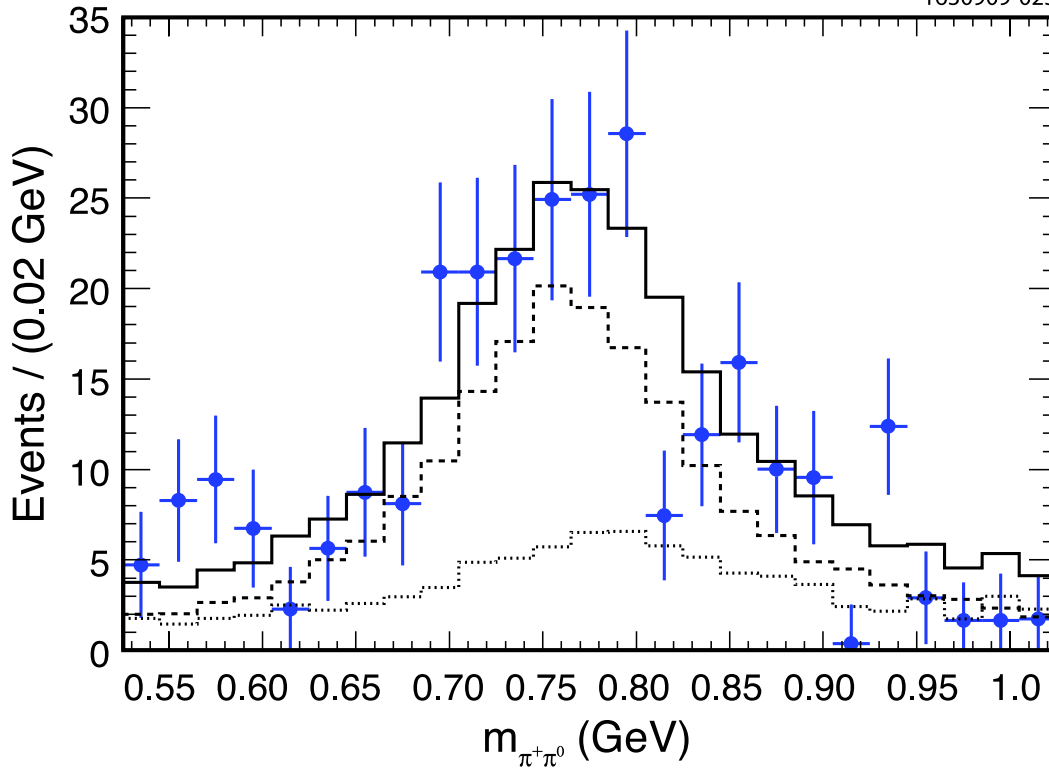}
\vspace{2mm}
\caption{Distribution of $\pi^+\pi^0$ mass.  The points with error bars are the sideband subtracted data. The dashed line represents the Monte Carlo predicted signal shape and the dotted line the Monte Carlo predicted background shape from real $D_s^+$ decays. The Monte Carlo predictions are normalized by the fitted yields to the data. The predicted total is given by the solid line. We require that  $E_{\rm extra}~<$ 0.2 GeV, and $ -0.05 ~< $MM$^2~ <$ 0.60 GeV$^2$.} \label{mass_rho}
\end{figure}

We also show the $\pi^+\pi^0$ mass distribution in Fig.~\ref{mass_rho}. Again the predicted sum has good agreement with the shape and data yield.

\section{Conclusions}

We list the CLEO-c  measurements of leptonic branching ratios and  $f_{D_s}$ in Table~\ref{tab:fDs}.
To extract the decay constant we use  $M_{D_s^+}=1.96849(34)$ GeV, a $D_s^+$  lifetime of 0.500(7) ps, and 1.77684(17) GeV for the $\tau^+$ mass \cite{PDG}. While it has been customary to take $|V_{cs}|= |V_{ud}|$, the expansion of the Wolfenstein parametrization of the CKM matrix to order $\lambda^4$ \cite{Charles} implies that $|V_{cs}|= |V_{ud}|-|V_{cb}|^2/2$.  We use  $|V_{ud}|=0.97418(26)$ as derived in Ref.~\cite{Towner-Hardy}. For $|V_{cb}|$ we use  a value of 0.04 from an average of exclusive and inclusive semileptonic $B$ decay results as discussed in Ref.~\cite{ABS}.  Thus, we find $|V_{cs}|= 0.97338(26)$. The resulting value of $f_{D_s}$ is 0.2 MeV larger than taking  $|V_{cs}|=|V_{ud}|$.

 Previously reported values of $f_{D_s}$ have been corrected to corresponded to the above numbers. These quantities contribute additional small amounts to the systematic error of  $\pm$1.8 MeV (lifetime), $\pm$0.1 MeV ($V_{cs}$), and for the $\tau^+\nu$ mode only $\pm$0.4 MeV ($M_{D_s^+}$) and $\pm0.2$ MeV ($\tau^+$ mass), that are included in the quoted values.
A theoretical upper bound on $f_{D_s}$ of 270 MeV has been calculated using two-point
correlation functions by Khodjamirian \cite{Kho}. The CLEO-c values for both the $\tau^+\nu$ and $\mu^+\nu$ modes are below this limit.
\begin{table}[htb]
\begin{center}
\caption{Recent absolute measurements of $f_{D_s}$ from CLEO-c\label{tab:fDs}}.
\begin{tabular}{lcccc} \hline\hline
Experiment & Mode &${\cal B}$ (\%) & $f_{D_s}$ (MeV)\\\hline
This result & $\tau^+\nu~(\rho^+\overline{\nu})$ &
$(5.52\pm 0.57\pm 0.21)$& $257.8\pm 13.3 \pm 5.2 $ \\
CLEO-c \cite{Dstomunu}& $\tau^+\nu,~(\pi^+\overline{\nu})$  & $(6.42\pm 0.81\pm 0.18) $& $278.0\pm 17.5 \pm 4.4 $ \\
CLEO-c  \cite{CLEO-CSP}& $\tau^+\nu~(e^+\nu\overline{\nu})$ &
$(5.30\pm 0.47\pm 0.22)$& $252.6\pm 11.2 \pm 5.6 $ \\\hline
Average & $\tau^+\nu$  &$(5.58\pm 0.33\pm 0.13)$& $259.7\pm 7.8\pm  3.4$\\\hline
CLEO-c  \cite{Dstomunu}& $\mu^+\nu$&
$(0.565\pm 0.045\pm 0.017)$ & $257.6\pm 10.3\pm 4.3$\\\hline
%Belle \cite{Belle-munu}
%& $\mu^+\nu$ & $(0.644\pm 0.076\pm 0.057)$& $275\pm 16\pm 12 $ \\\hline
%Average & $\mu^+\nu$  &$(0.581\pm 0.039\pm 0.016)$& $261.1 \pm  8.8 \pm  3.6$\\\hline
Average & $\tau^+\nu$+$\mu^+\nu$& & $259.0 \pm  6.2 \pm 3.0$\\
 \hline\hline
\end{tabular}
\end{center}
\end{table}

The ratio of decay constants from CLEO for the two leptonic decay modes is
\begin{equation}
\frac{f_{D_s}\left(D_s^+\to\tau^+\nu\right)}{f_{D_s}\left(D_s^+\to\mu^+\nu\right)}=1.01\pm 0.05,
\end{equation}
consistent with lepton universality.
The average value of the pseudoscalar decay constant using  both  leptonic decay modes is
\begin{equation}
f_{D_s} =(259.0 \pm 6.2 \pm3.0){\rm~ MeV}.
\end{equation}

There are two SM based theoretical predictions for $f_{D_s}$ in the literature based on Lattice QCD calculations, where all
three light quark loops are included. The values predicted are ($241\pm 3$) MeV from the HPQCD+UKQCD collaboration \cite{Lat:Foll}, and ($249\pm 11$) MeV from the FNAL+MILC+HPQCD collaboration \cite{Lat:Milc}.  We choose to compare with the more precise Follana \etal~result, realizing that it needs confirmation, especially with respect to the rather small error.
The difference between the experimental average of $f_{D_s}$ and the HPQCD+UKQCD prediction is 2.4  standard deviations. Other theoretical predictions are given in Ref.~\cite{Rosner-Stone}.

 Belle has also measured the absolute branching fraction for $D_s^+\to \mu^+\nu$ and found a value  $f_{D_s}=(275\pm 16\pm 12 )$ MeV \cite{Belle-munu,Otherfds}. Combining with the CLEO measurements we find $f_{D_s}=(260.7 \pm  6.5$) MeV, which differs from the HPQCD+UKQCD prediction by 2.8 standard deviations.  We emphasize that this difference is qualitatively different than looking for new physics as a bump in mass spectrum where any values of the mass and width can be entertained. Here we are dealing with a theoretical number that was predicted before the measurements were available.  Thus, although we cannot claim to have definitely seen an important discrepancy pointing to physics beyond the standard model, it is important to pay attention to this difference and to see what it may imply.

In fact this possible discrepancy has motivated several new beyond
the SM theories. These include leptoquark models of Dobrescu and Kronfeld \cite{Dobrescu-Kronfeld},
R parity violating models of  Akeroyd and Recksiegel \cite{Rviolating}, and Kundu and Nandi  who relate this discrepancy with preliminary indications of a large phase in $B_s-\overline{B}_s$ mixing, and explain both with a specific supersymmetry model \cite{KN}. Dosner \etal~\cite{Dosner} show however, that scalar leptoquark and R-parity violating models would have different effects on $\tau^+\nu$ and $\mu^+\nu$ final states.
Gninenko and Gorbunov
argue that the neutrino in the $D_s$ decay mixes with a sterile neutrino, which
enhances the rate and also explains the excess number of low energy electron
like events in the MiniBooNE data \cite{GG}.

We also have measured the following branching fractions:
\begin{eqnarray}
{\cal B}(D_s^+\to K^0\pi^+\pi^0)&=&(1.00\pm0.18\pm 0.04)\%,\\\nonumber
{\cal B}(D_s^+\to\pi^+\pi^0\pi^0)&=&(0.65\pm0.13\pm 0.03)\%,\\\nonumber
{\cal B}(D_s^+\to\eta\rho^+)&=&(8.9\pm0.6\pm0.5)\%.
\end{eqnarray}
The first two modes have not been measured previously.

\section{Acknowledgments}
We thank Nabil Menaa for useful discussions.
We gratefully acknowledge the effort of the CESR staff
in providing us with excellent luminosity and running conditions.
D.~Cronin-Hennessy and A.~Ryd thank the A.P.~Sloan Foundation.
This work was supported by the National Science Foundation,
the U.S. Department of Energy,
the Natural Sciences and Engineering Research Council of Canada, and
the U.K. Science and Technology Facilities Council.

\section*{Appendix A: Monte Carlo Generated Background Modes }
We list the different modes that populate the MM$^2$ distribution for the three $E_{\rm extra}$ intervals as given by the generic Monte Carlo simulation in Table~\ref{tab:bkgrd}.
\begin{table}[htb]
\begin{center}
\caption{Assumed branching fractions, Numbers of events (\#) and fractions resulting from a generic Monte Carlo simulation of the $D_s^+$ backgrounds for twenty times the data in the interval $-0.2 <$ MM$^2< 1.0$ GeV$^2$. \label{tab:bkgrd}}
\begin{tabular}{ccrcrcrc} \hline\hline
Mode&${\cal B}$(\%)&\multicolumn{2}{c}{$0<E_{\rm extra}<0.1$ GeV} & \multicolumn{2}{c}{$0.1<E_{\rm extra}<0.2$ GeV} &\multicolumn{2}{c}{0.8 GeV $<E_{\rm extra}$} \\
 & &\#~ & Fraction (\%) & \#~ & Fraction (\%)& \#~ & Fraction (\%) \\\hline
$K^0\pi^+\pi^0$& 0.85& 348 & 29.3 & 128 & 10.3 & 548 & 1.61\\
$\eta\rho^+$ & 7.58&114&9.6 & 215 & 17.4 & 17797 & 52.4\\
$\pi^+\pi^0\pi^0$&0.58&48&4.1&27&2.2&991&2.9\\
$\tau^+\nu$, $\tau^+\to\pi^+\pi^0\pi^0\bar{\nu}$&0.55&159&13.4&266&21.5&99 &0.3\\
$\tau^+\nu$, $\tau^+\to\pi^+\bar{\nu}$&0.66&63&5.3&18&1.5&2 &0.01\\
$\tau^+\nu$, $\tau^+\to$ other&3.27& 81&6.8&43&3.5&12&0.04\\
$\phi\pi^+$, $\phi\to K_L^0K_S^0$&1.38&70&5.9&114&9.2&560&1.7\\
$X\mu^+\nu$&5.87 &91&7.7&158&12.8&653&1.9\\
$\eta\pi^+$&1.54&15&1.3&32&2.6&1250&3.7\\
$\mu^+\nu$&0.61&19&1.6&15&1.2&9&0.03\\
$\eta'\pi^+$&3.67&10&0.84&15&1.2&2610&7.7\\ %\hline
$K^0K^+\pi^0$&2.50&20&1.7&7&0.6&32&0.09\\
$K^0K^+$&2.93&11&0.9&6&0.5&31&0.09\\
$K^0\pi^+$&0.24&9&0.8&9&0.7&59&0.17\\
$K_S^0K_S^0\pi^+$&0.70&0&0&1&0.1&156&0.46\\
$K_S^0K_L^0\pi^+$&1.25&29&2.5&62&5.0&144&0.42\\
$K_L^0K_L^0\pi^+$&0.70&18&1.5&8&0.7&31&0.09\\
$Xe^+\nu$&6.19&8&0.7&5&0.4&16&0.05\\
$K^0\pi^+\pi^0\pi^0$&0.65&24&2.0&47&3.8&457&1.34\\
$\eta\pi^+\pi^0\pi^0$&3.25&5&0.4&12&1.0&4357&12.82\\
$\eta'\pi^+\pi^0$&3.87&2&0.2&6&0.5&2041&6.00\\
$\omega\pi^+$&0.25&1&0.1&0&0&18&0.05\\
$\pi^+\pi^0\pi^0\pi^0\pi^0$&0.85&1&0.1&3&0.2&1684&4.95\\
$\phi\pi^+\pi^0$&7.35&4&0.3&14&1.1&340&1.00\\
Other&&36&3.0&27&2.2&97&0.29\\
\hline\hline
\end{tabular}
\end{center}
\end{table}

\section*{\boldmath Appendix B: Measurements of  $D_s^+$ Branching Fractions for Selected Background Modes}

\begin{figure}[hbt]
\centering
\includegraphics[width=4in]{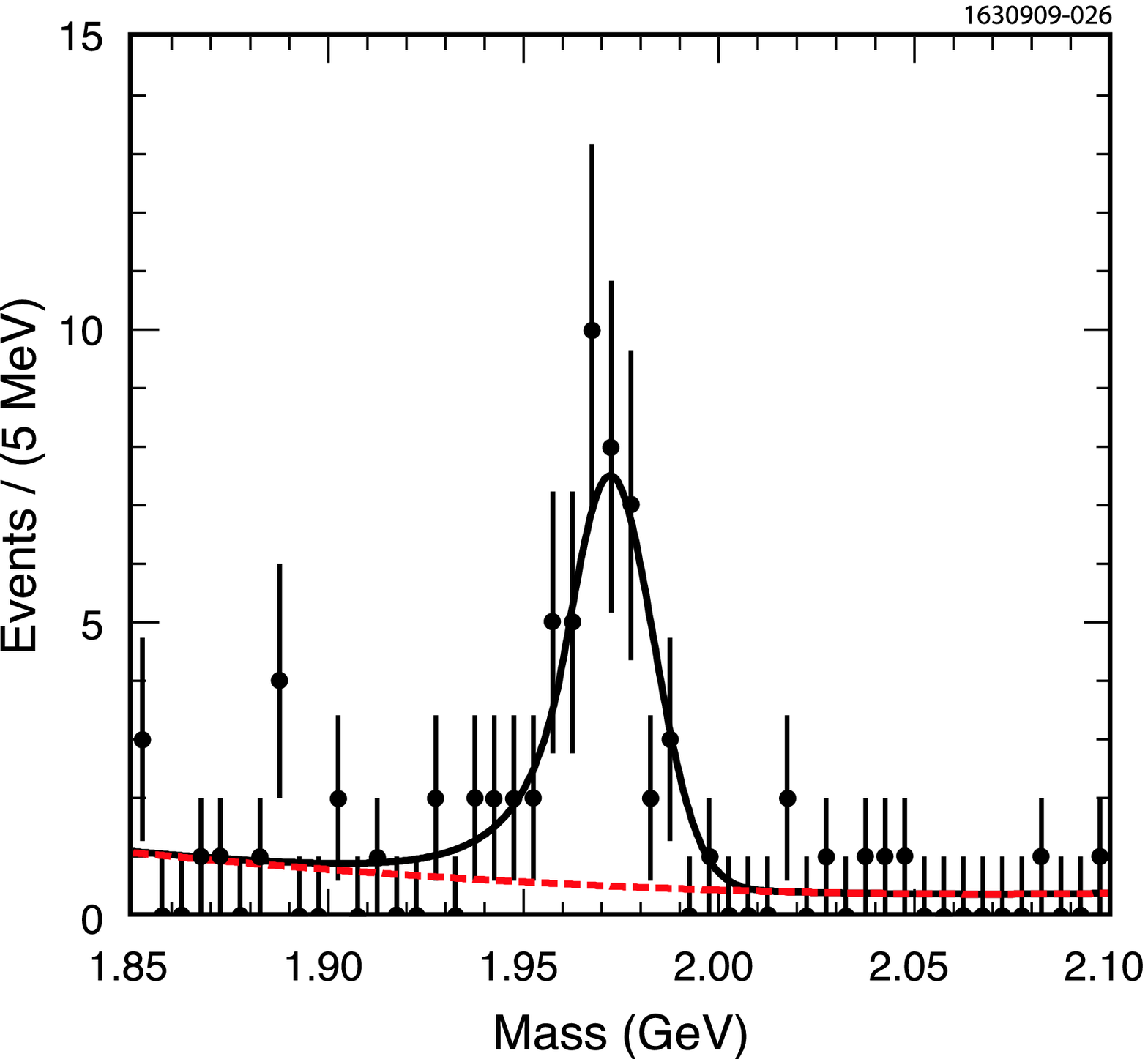}
\vspace{2mm}
\caption{The invariant mass spectrum of $K_S^0\pi^+\pi^0$. The curves show a second order Chebyshev polynomial function that describes the background summed with a signal CB function whose width is fixed (solid line).
} \label{kspipi0}
\end{figure}

The $K^0\pi^+\pi^0$ mode has not been previously measured.  We select events opposite our $D_s^-$ tag candidates with a single charged track consistent with being a $\pi^+$ in conjunction with a $\pi^0$ candidate as described above, and a $K_S^0\to\pi^+\pi^-$ candidate where the invariant $\pi^+\pi^-$ mass is within 12 MeV of the known $K_S^0$ mass and the flight significance, the distance that the $K_S$ travels divided by the error in the distance,  is greater than 2.  The invariant mass of  $K_S^0\pi^+\pi^0$ combinations is shown in Fig.~\ref{kspipi0}. The data are fit with a signal  CB function with all parameters except the area fixed to those given by the Monte Carlo simulation of this mode, and a second order Chebyshev background polynomial. The fit yields 44$\pm$8 events.

\begin{figure}[hbt]
\centering
\includegraphics[width=4in]{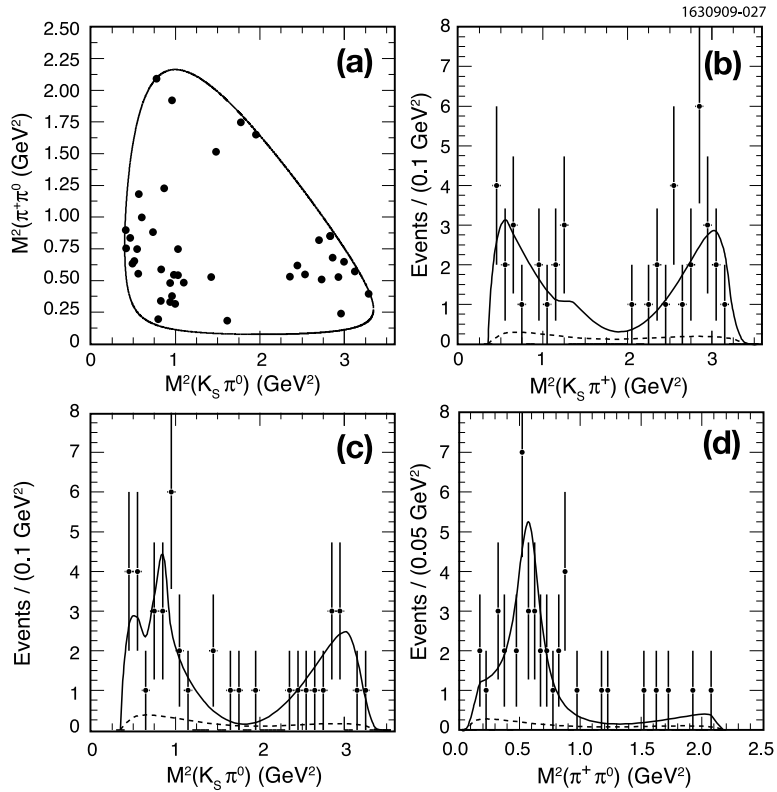}
\vspace{2mm}
\caption{(a) Dalitz plot of $K_S^0\pi^+\pi^0$, where the invariant mass of the three particles is selected within $\pm$20 MeV of the $D_s^+$ mass.  (b)--(d) show the mass projections, the solid curves show the overall fit and the dashed curve the background
from sidebands.
 } \label{d_kspipi0}
\end{figure}

We proceed by performing an unbinned liklihood fit to the Daltiz plot  shown in Fig.~\ref{d_kspipi0}, using the isobar model formalism as described in Ref.~\cite{isobar}. The fit results are that the
$K_S^0\rho^+$ fraction is (88$\pm$8)\% with only (17$\pm$8)\% of $K^{*+}\pi^0$; the relative phase is (21$\pm$25) degrees.  The efficiency is determined to be 21\% by Monte Carlo simulation which uses a $K\rho$ resonance structure, resulting in a branching fraction
\begin{equation}
{\cal B}(D_s^+\to K^0\pi^+\pi^0)=(1.00\pm0.18\pm 0.04)\%,
\end{equation}
where the systematic error arises from several sources shown in Table~\ref{tab:sysbk}. The error due to the Dalitz plot structure is  is 0.7\%, found by evaluating the relative efficiency difference between pure $K_S\rho^+$ and the model resulting from our Dalitz plot fit; this is negligible compared to the other sources.  Note that for our purposes some of the systematic error cancels because we are using the same $D_s^-$ tag sample and the same $\pi^+$ and $\pi^0$ detection efficiencies as for signal $\tau^+\to\rho^+\overline{\nu}$.

\begin{table}[htb]
\begin{center}
\caption{Systematic errors on determination of the branching fractions of several
background modes. We give two errors, one for the branching fraction to be used as an independent measurement (Ext), and the second is the one to be used internally (Int) for the lepton branching fraction analysis where some of the errors cancel.\label{tab:sysbk}}
\begin{tabular}{lcccccc} \hline\hline
&\multicolumn{2}{c}{$D_s^+\to K^0\pi^+\pi^0$}&\multicolumn{2}{c}{$D_s^+\to \eta\rho^+$}
&\multicolumn{2}{c}{$D_s^+\to \pi^+\pi^0\pi^0$}\\
Error Source &(Ext)(\%) & (Int)(\%)&(Ext)(\%) & (Int)(\%)&(Ext)(\%) & (Int)(\%) \\ \hline
Hadron identification & 1.0 &0 & 1.0 &0& 1.0 &0\\
Finding $\pi^0$  from ($\rho^+$) &1.3  & 0&1.3  & 0&1.3  & 0\\
Background modeling& 2.0 & 2.0 & 1.9 & 1.9 & 2.0 & 2.0\\
$\pi^0$ efficiency & 1.3 & 0& 1.3 & 0& 2.6 & 1.3\\
$K^0$ efficiency &2.0 & 2.0 & 0 &0 & 0 & 0\\
$\eta$ efficiency& 0 & 0 & 4.0 & 4.0 & 0 & 0\\
Number of tags& 2.0 & 0& 2.0 & 0& 2.0 & 0\\
Tag bias & 1.0 & 0& 1.0 & 0& 1.0 & 0 \\\hline
Total & 3.8 & 2.8 & 5.1 & 4.4 & 4.0 & 2.4\\
\hline\hline
\end{tabular}
\end{center}
\end{table}

\begin{figure}[hbt]
\centering
\includegraphics[width=4in]{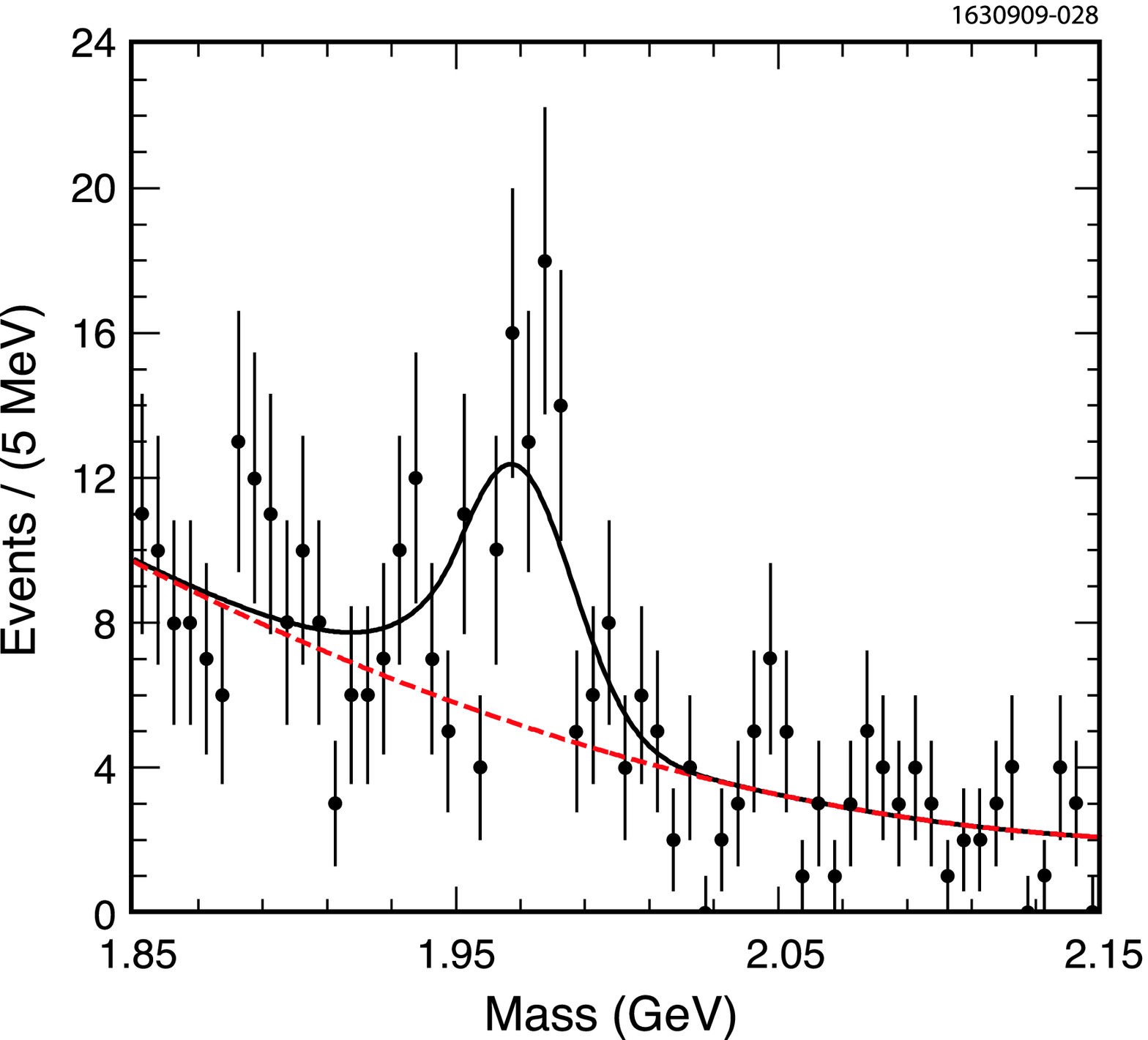}
\vspace{2mm}
\caption{The invariant mass spectrum of $\pi^+\pi^0\pi^0$ candidates. The curves show a second order Chebyshev polynomial function that describes the background (dashed line) summed with signal CB function (solid line) whose width is fixed to the Monte Carlo predicted shape plus the measured Gaussian smearing.
 } \label{pipi0pi0}
\end{figure}

The $\pi^+\pi^0\pi^0$ mode also has not previously been measured, though the analogous isospin related mode $\pi^+\pi^+\pi^-$ has been. Here we require that the invariant $\pi^0\pi^0$ mass be more than 50 MeV from the $K_S^0$ mass in order to reject $K_S^0$.
The invariant mass plot is shown in Fig.~\ref{pipi0pi0}. The signal CB function is fixed to the Monte Carlo predicted shape, and we use a second order Chebyshev polynomial function to model the background shape. We find a signal of $72\pm 16$ events.

\begin{figure}[hbt]
\centering
\includegraphics[width=4in]{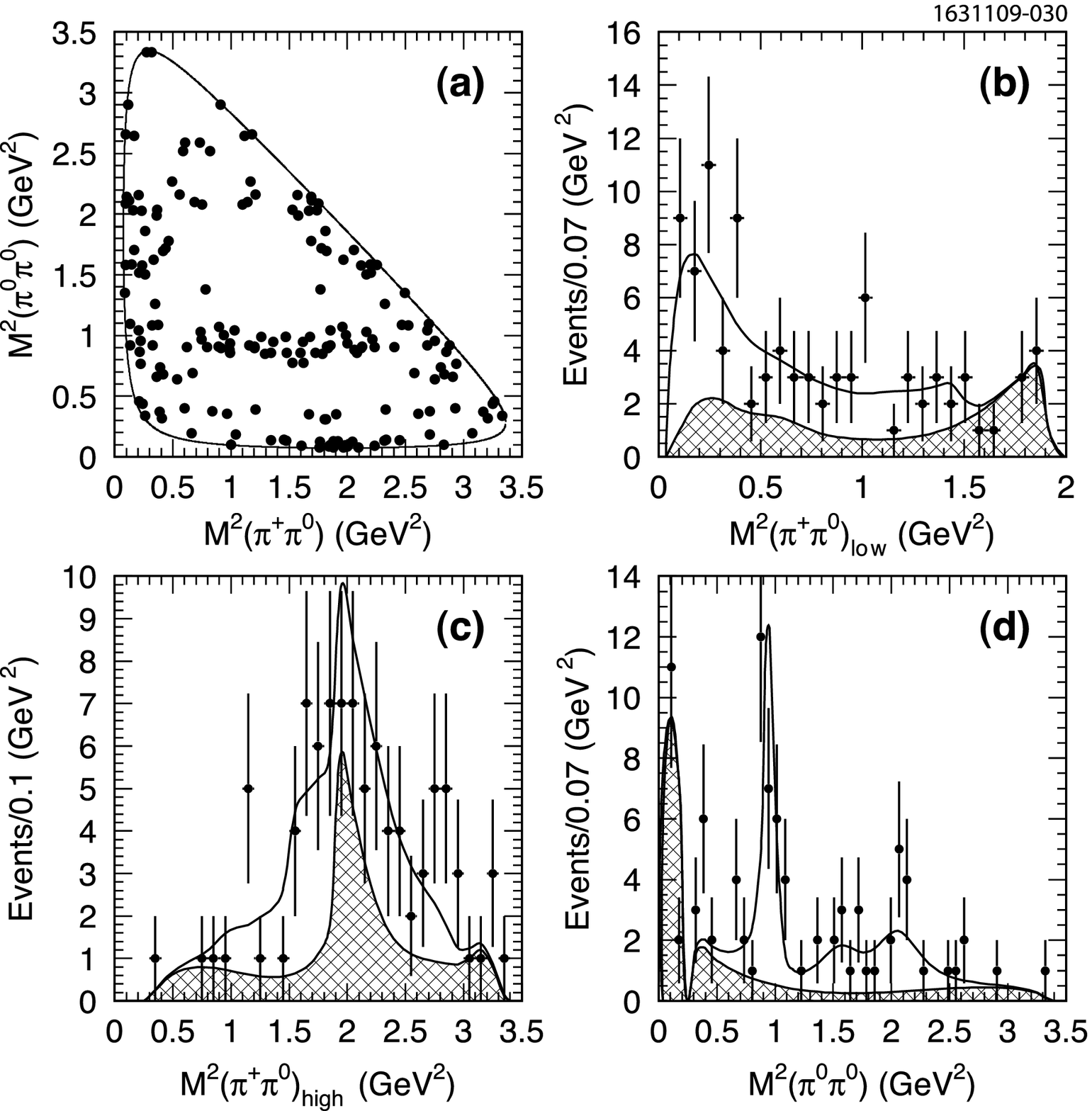}
\vspace{2mm}
\caption{(a) Dalitz plot of $\pi^+\pi^0\pi^0$, where the invariant mass of the three particles is selected within $\pm$24 MeV of the $D_s^+$ mass. There are two entries per event.  (b)--(d) show the mass projections, the solid curves show the overall fit and the shaded region the background from sidebands. The zero at 0.25 GeV$^2$ in (d) results from the $K_S$ rejection criteria.
 } \label{d_pipipi0}
\end{figure}

We also perform an unbinned liklihood fit to the Daltiz plot  shown in Fig.~\ref{d_pipipi0},  again using the isobar model formalism. There is no evidence for $\rho^+\pi^0$. We find that the fractions of
$f_0(980)\pi^+$, $f_2(1270)\pi^+$ and $f_0(1370)\pi^+$ are (56.5$\pm$9.1)\%,  (20.5$\pm$7.3)\%, and (38.1$\pm$8.6)\%, respectively. Fixing the  $f_0(980)\pi^+$ phase at zero degrees, the relative phases of the $f_2(1270)\pi^+$ and $f_0(1370)\pi^+$ with respect to zero are (243$\pm$29) degrees and (299$\pm$24) degrees, respectively.

The Monte Carlo simulated efficiency is 28.1\%, yielding
\begin{equation}
{\cal B}(D_s^+\to\pi^+\pi^0\pi^0)=(0.65\pm0.13\pm 0.03)\%.
\end{equation}
The systematic errors are listed in Table~\ref{tab:sysbk}. This number is consistent with ${\cal B}(\pi^+\pi^+\pi^-)/2=(0.56\pm0.04)$\% \cite{PDG}, which is what is expected if  the neutral dipion system final state dominates both modes.

\begin{figure}[hbt]
\centering
\includegraphics[width=4in]{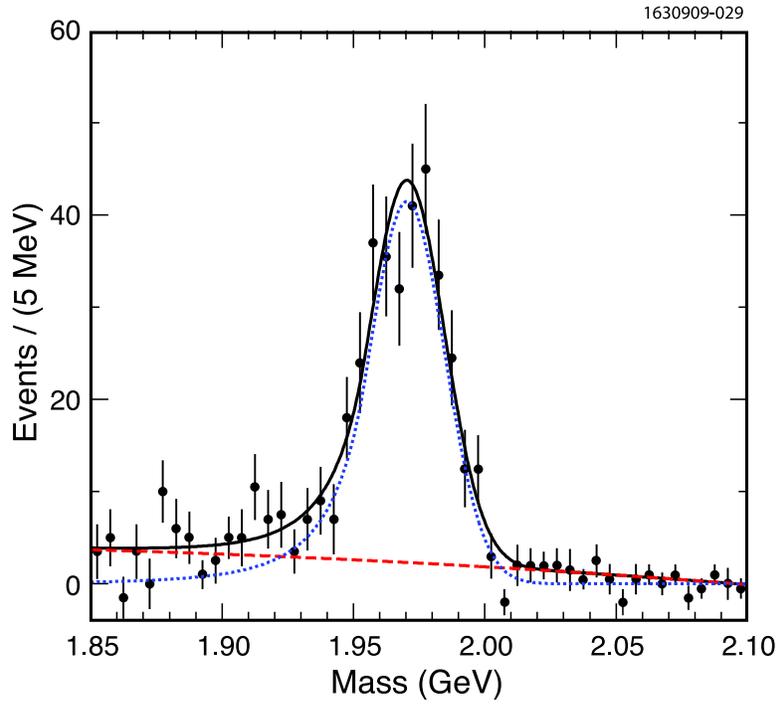}
\vspace{2mm}
\caption{The invariant mass spectrum of $\eta\pi^+\pi^0$ candidates, where $\eta\to\gamma\gamma$. The curves show a signal CB function whose width is allowed to float (dotted line), a second order Chebyshev polynomial function that describes the background (dashed line), and the sum (solid line).
 } \label{etarho}
\end{figure}

The $\eta\rho^+$ branching fraction has been previously measured as ($13.1\pm2.2)$\% \cite{PDG}; we wish to improve on this accuracy. We look for events with only one  charged track consistent with being a $\pi^+$ The $\eta$ is looked for in the $\gamma\gamma$ decay mode only; mass combinations are used if they are within 3 standard deviations of the $\eta$ mass. We insist that the $\pi^+\pi^0$ invariant mass be within 250 MeV of the $\rho^+$ mass. The resulting $\gamma\gamma\pi^+\pi^0$ invariant mass spectrum is shown in Fig.~\ref{etarho}. Here we have enough data to let the r.m.s.\ width of the CB function vary in the fit. The background is again described by a second order Chebyshev function. We find a total of $328\pm22$ events. We use a Monte Carlo determined efficiency of 22.4\%. We find that
\begin{equation}
{\cal B}(D_s^+\to\eta\rho^+)=(8.9\pm0.6\pm0.5)\%.
\end{equation}
The systematic errors are listed in Table~\ref{tab:sysbk}.  Our new measurement is lower by about 1.8 standard deviations than the PDG average \cite{PDG}.

\afterpage{\clearpage}


\begin{thebibliography}{99}
\bibitem{Otherfds}
We have not included other measurements of $f_{D_s}$ in the average that are based on measuring the
leptonic branching fractions with respect to a hadronic decay mode due to the relatively large uncertainities
in the determination of the relevant hadronic branching fraction. See Ref.~\cite{Rosner-Stone} for a compilation
of these results.
\end{thebibliography}
\end{document}